\documentclass[onecolumn,conference,letterpaper]{IEEEtran}
\IEEEoverridecommandlockouts
\usepackage{cite}
\usepackage[cmex10]{amsmath}
\usepackage{amssymb,amsfonts}
\usepackage{bbm}
\usepackage{manfnt}
\usepackage{xfrac}
\usepackage{etoolbox}
\usepackage{booktabs}
\makeatletter
\patchcmd{\@opargbegintheorem}{(#3)}{#3}{}{}
\makeatother
\usepackage[inline]{enumitem}
\usepackage{subcaption}
\usepackage{standalone}
\usepackage{algorithmic}
\usepackage{graphicx}
\usepackage{textcomp}
\usepackage{xspace}
\usepackage{hyperref}
\providecommand{\texorpdfstring}[2]{#1}
\usepackage{url}
\usepackage{aliascnt}
\usepackage[capitalize,nameinlink]{cleveref}
\usepackage[dvipsnames]{xcolor}
\usepackage{ifthen}
\usepackage{tikz}
\usetikzlibrary{%
  arrows.meta,%
  automata,%
  backgrounds,%
  calc,%
  colorbrewer,%
  decorations.markings,%
  decorations.pathmorphing,%
  decorations.pathreplacing,%
  decorations.shapes,%
  decorations.text,%
  fit,%
  graphs,%
  graphs.standard,%
  hobby,%
  matrix,%
  shapes.geometric,%
  patterns,%
  positioning,%
  quotes,%
  tikzmark,%
  shadows,%
}
\usepackage{microtype}
\hypersetup{
    colorlinks = true,
    linkcolor = RubineRed,
    anchorcolor = RubineRed,
    citecolor = RubineRed,
    filecolor = RubineRed,
    urlcolor = RubineRed
}

\providecommand{\texorpdfstring}[2]{#1}
\newtheorem{theorem}{Theorem}
\newcommand{\NewThm}[3]{%
    \newaliascnt{#1}{theorem}
    \newtheorem{#1}[#1]{#2}
    \aliascntresetthe{#1}
    \crefname{#1}{#2}{#3}
}
\NewThm{lemma}{Lemma}{Lemmas}
\NewThm{prop}{Proposition}{Propositions}
\NewThm{corollary}{Corollary}{Corollaries}
\NewThm{conjecture}{Conjecture}{Conjectures}

\crefname{definition}{Definition}{Definitions}

\crefname{remark}{Remark}{Remarks}

\crefname{example}{Example}{Examples}

\crefformat{section}{\S#2#1#3}
\crefrangeformat{section}{\S#3#1#4 to \S#5#2#6}
\crefmultiformat{section}{\S#2#1#3}%
{ and~\S#2#1#3}{, \S#2#1#3}{ and~\S#2#1#3}

\newcommand{\V}[1]{\mathbf{#1}}
\newcommand{\field}[1][q]{\mathbb{F}_{#1}}
\newcommand{\Field}[1][q]{\mathbb{F}_{#1}}
\newcommand{\msg}{\V{m}}
\newcommand\xqed[1]{%
  \leavevmode\unskip\penalty9999 \hbox{}\nobreak\hfill
  \quad\hbox{#1}}
\newcommand{\triqed}{\xqed{\mantriangleright}}

\newcommand{\node}{symbol\xspace}

\newcommand{\piggyfw}{Piggybacking framework\xspace}

\newcommand{\Code}{\mathcal{C}}
\newcommand{\Initial}{I} 
\newcommand{\Final}{F} 
\newcommand{\ICode}{{\Code^{\Initial}}} 
\newcommand{\FCode}{{\Code^{\Final}}}

\newcommand{\In}{n^{\Initial}} 
\newcommand{\Ik}{k^{\Initial}} 
\newcommand{\Ir}{r^{\Initial}}

\newcommand{\Is}{\lambda^{\Initial}}

\newcommand{\Fn}{n^{\Final}} 
\newcommand{\Fk}{k^{\Final}} 
\newcommand{\Fr}{{r^{\Final}}}

\newcommand{\Fs}{\lambda^{\Final}}

\newcommand{\Msg}{\V{m}}

\newcommand{\sourcev}{\textsf{\textbf{s}}}
\newcommand{\sinkvs}[1]{\textsf{\textbf{t}}_{#1}}
\newcommand{\sinkv}{\textsf{\textbf{t}}}
\newcommand{\centralv}{\textsf{\textbf{c}}}

\newcommand{\SplitCode}{$(\In, \Ik = \Fs\Fk;\allowbreak \Fn, \Fk)$ convertible code\xspace}

\newcommand{\scalingall}{zhang2010alv,zheng2011fastscale,wu2012gsr,zhang2014rethinking,huang2015scale,wu2016i/o,zhang2018optimal,hu2018generalized,zhang2020efficient,rai2015adaptivea,rai2015adaptive,wu2020optimal,hu2021ncscale}



\newenvironment{proof}{\begin{IEEEproof}}{\end{IEEEproof}}

\AtEndEnvironment{example}{\triqed}

\begin{document}

\title{Bandwidth Cost of Code Conversions\\in the Split Regime
}

\author{%
\IEEEauthorblockN{Francisco Maturana and K. V. Rashmi}
\IEEEauthorblockA{%
Carnegie Mellon University, Pittsburgh, PA, USA \\
Email: fmaturan@cs.cmu.edu, rvinayak@cs.cmu.edu}%
}

\maketitle
\thispagestyle{plain}
\pagestyle{plain}

\begin{abstract}
Distributed storage systems must store large amounts of data over long periods of time.
To avoid data loss due to device failures, an $[n,k]$ erasure code is used to encode $k$ data symbols into a codeword of $n$ symbols that are stored across different devices.
However, device failure rates change throughout the life of the data, and tuning $n$ and $k$ according to these changes has been shown to save significant storage space.
Code conversion is the process of converting multiple codewords of an initial $[\In,\Ik]$ code into codewords of a final $[\Fn,\Fk]$ code that decode to the same set of data symbols.
In this paper, we study \emph{conversion bandwidth}, defined as the total amount of data transferred between nodes during conversion.
In particular, we consider the case where the initial and final codes are MDS and a single initial codeword is split into several final codewords ($\Ik=\Fs\Fk$ for integer $\Fs \geq 2$), called the \emph{split regime}.
We derive lower bounds on the conversion bandwidth in the split regime and propose constructions that significantly reduce conversion bandwidth and are optimal for certain parameters.
\end{abstract}

\begin{figure}
    \centering
    \includegraphics[width=.7\linewidth]{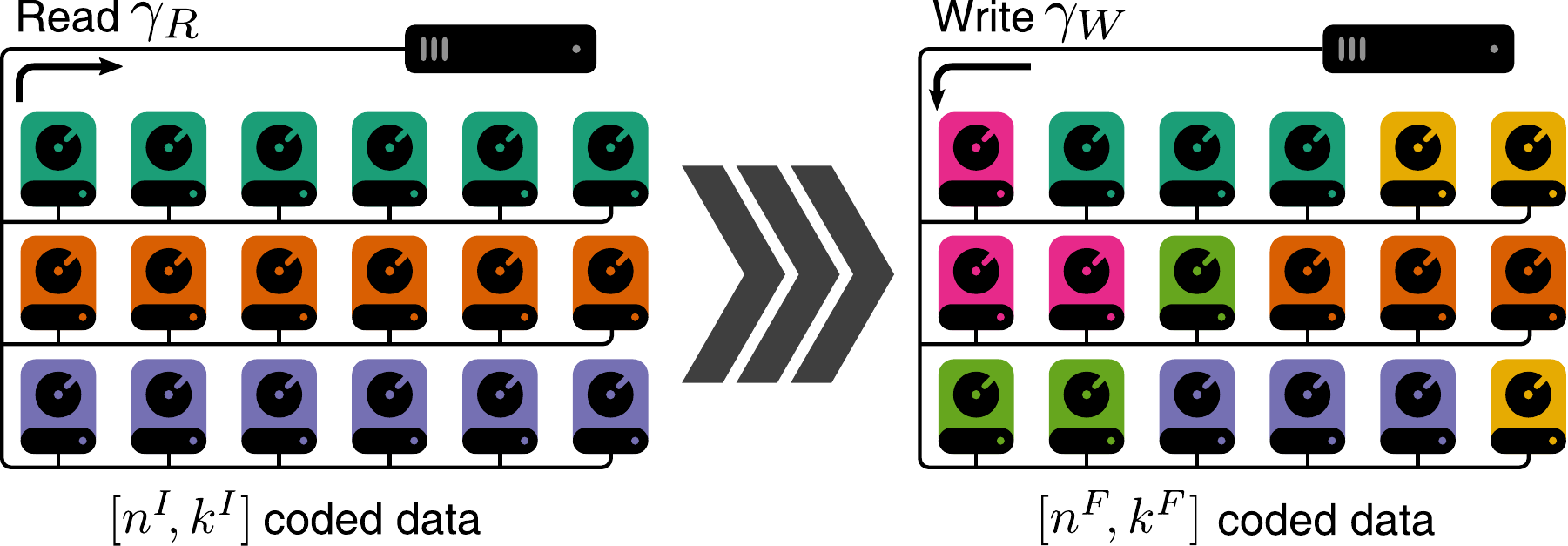}
    \caption{Example of code conversion from code $[\In,\Ik]$ to code $[\Fn,\Fk]$. Each color denotes a different codeword. Conversion bandwidth is defined as the total amount of data read or written during conversion, i.e.\ ($\gamma_R + \gamma_W$).} \label{fig:conversion-diagram}
\end{figure}

\section{Introduction}
Distributed storage systems use erasure codes to store large amounts of data reliably and without excessive storage overhead~\cite{ghemawat2003google,borthakurhdfs,huang2012erasure,asfapache}.
An $[n, k]$ erasure code encodes $k$ symbols of data into a codeword with $n$ symbols, which are then stored in different storage devices.
If the code is maximum-distance-separable (MDS), then the full data can be decoded even after $n-k$ concurrent device failures.

Data in distributed storage systems is usually stored over long periods of time.
Kadekodi et al.~\cite{kadekodi2019cluster} showed that the failure rate of devices can significantly change over this time and that tuning the parameters $n$ and $k$ to adjust to these changes results in significant savings in storage space.
In most cases, this tuning requires changing $n$ and $k$ simultaneously due to practical system constraints~\cite{kadekodi2019cluster}.
Other reasons to change $n$ and $k$ include adapting to changes in data popularity or space availability.
Whenever $n$ and $k$ are changed, all the data that is already encoded must be modified to conform to the newly chosen parameters.
The \emph{default approach} to performing this change is to read all the data (decoding if necessary), re-encode with the new $n$ and $k$, and write back to the storage devices.
This results in very high consumption of cluster resources~\cite{kadekodi2019cluster}, such as network bandwidth, IO, and CPU, which can overwhelm the cluster for periods of several days.

The \textit{code conversion} problem, introduced in~\cite{maturana2020convertible}, provides a theoretical framework to study the problem of efficiently changing the code parameters for already encoded data.
Code conversion is the process of changing multiple (already encoded) codewords of an initial code of parameters $[\In, \Ik]$ to multiple codewords of a final code of parameters $[\Fn,\Fk]$ (\cref{fig:conversion-diagram}).
Let $\Ir := \In - \Ik$ and $\Fr := \Fn - \Fk$.
The main goal of the study of code conversion~\cite{maturana2020convertible,maturana2020access,maturana2021bandwidth} is to design the initial and final codes, as well as a conversion procedure, which can convert encoded data more efficiently than the default approach, for given parameters $(\In,\Ik;\Fn,\Fk)$.
Codes designed for this purpose are referred to as \emph{convertible codes}.
The initial work on convertible codes~\cite{maturana2020convertible,maturana2020access} addressed this challenge by focusing on the \emph{access cost} of conversion, defined as the number of code symbols that are either read or written during conversion.
In~\cite{maturana2020convertible,maturana2020access}, the authors showed that access cost can be significantly reduced compared to the default approach.

In~\cite{maturana2021bandwidth}, the authors introduced convertible codes optimized for another important metric: network bandwidth.
Here, the cost of conversion is measured in terms of \emph{conversion bandwidth}, defined as the total amount of data transferred between nodes during conversion, which is divided into read conversion bandwidth ($\gamma_R$) and write conversion bandwidth ($\gamma_W$).
The work~\cite{maturana2021bandwidth} focused exclusively on a parameter regime known as the \emph{merge regime}, which consists of conversions that merge multiple codewords together (i.e.\ $\Fk = \Is\Ik$ for integer $\Is \geq 2$), and showed that conversion bandwidth can be significantly reduced compared to both the default approach and the codes that optimize the access cost of conversions.

\begin{figure}[t]
    \centering
    \vspace{-1.5em}
    \makebox[\textwidth][c]{
    \includegraphics[width=1.04\textwidth]{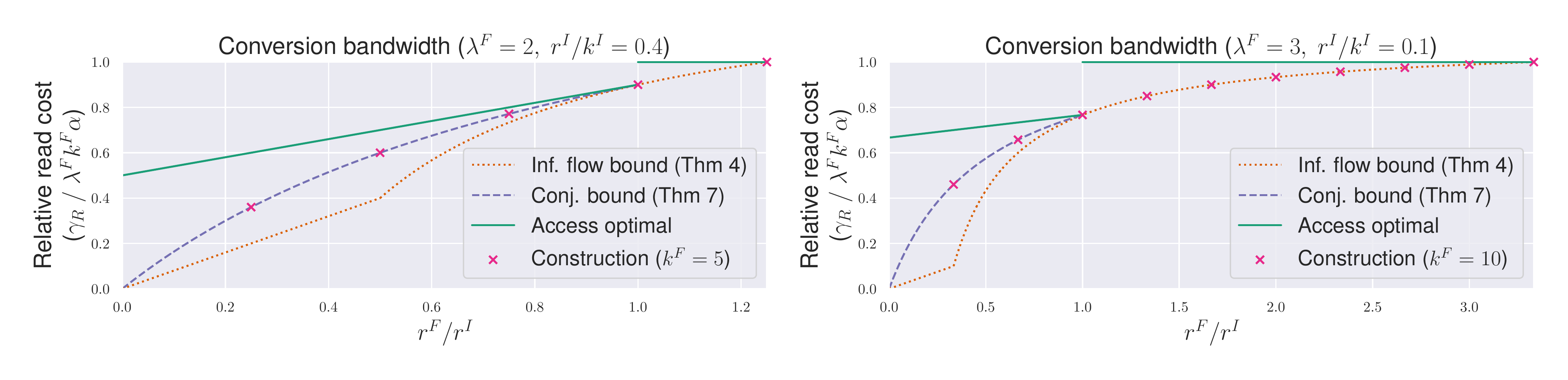}
    }
    \vspace{-3em}
    \caption{%
    Read-conversion-bandwidth relative to the default approach (split regime).
    In each plot, the value of the parameters $\Fs$ and the ratio $\Ir/\Ik$ are fixed, and the value of the ratio $\Fr/\Ir$ ranges in $(0, (\Fs\Ir/\Ik)^{-1}]$.
    By choosing this parametrization, the plotted curves are independent of the value of $\Fk$.
    For illustration, markers are added on the points that can be achieved by the construction of \cref{sec:split-construction} when $\Fk$ takes the given example value.
    }
    \label{fig:split-bandwidth-plot}
\end{figure}

In this paper, we study optimizing the conversion bandwidth for another important regime called the \emph{split regime}, wherein a single initial codeword is split into final codewords, i.e.\ $\Ik = \Fs\Fk$ for some integer $\Fs \geq 2$. 
In particular, we derive lower bounds on the conversion bandwidth of codes in the split regime, and we propose constructions that match those bounds.
The split regime is important because it plays a key role in the conversions with arbitrary parameters $(\In,\Ik;\Fn,\Fk)$~\cite{maturana2020access}.

We first focus on lower bounding conversion bandwidth in the split regime.
To do this, we model conversion using an information flow graph with edges of \emph{variable capacities}.
Using this model, we derive a lower bound on conversion bandwidth for convertible codes in the split regime satisfying some technical conditions (\cref{thm:split-bound-loose}).
This bound shows that savings are not possible when $\Fn \geq 2\Fk$ but leaves room for significant savings otherwise.
However, we show that this bound is not tight.
For this reason, we introduce a conjecture (\cref{thm:conjecture}) about the relationship between the amount of data that needs to be downloaded from different types of code symbols.
Assuming this conjecture is true, we derive an additional lower bound (\cref{thm:split-bound-tight}).
Finally, we present constructions which achieve the combination of both lower bounds (\cref{thm:split-construction}).
When $\Ir \leq \Fr$, these constructions achieve the lower bound of \cref{thm:split-bound-loose}, i.e.\ the optimal conversion bandwidth.

The proposed constructions can perform conversion with significantly less conversion bandwidth compared to the default approach.
Moreover, these constructions also require less conversion bandwidth than existing convertible codes optimized for access cost~\cite{maturana2020access}.
\Cref{tab:bw-comparison} and \cref{fig:split-bandwidth-plot} compare the read conversion bandwidth of the different approaches (we omit write conversion bandwidth because it is the same for all three approaches).

We start by summarizing the relevant background and related work in \cref{sec:background}.
Then, we model conversion in the split regime and derive lower bounds for conversion bandwidth in \cref{sec:split-lower-bound}.
Finally, we present a construction matching those lower bounds in \cref{sec:split-construction}.

\begin{table}
    \vspace{2em}
    \renewcommand{\arraystretch}{1.1}
    \centering
    \caption{Comparison of the read conversion bandwidth (read BW) of different approaches for split conversion.}
    \label{tab:bw-comparison}
    \begin{tabular}{ccc}
        \toprule
        Approach & Read BW ($\Ir < \Fr$) & Read BW ($\Ir \geq \Fr$) \\\midrule
        Default 
        & $\Fs\Fk\alpha$
        & $\Fs\Fk\alpha$ \\
        Access optimal \cite{maturana2020access}
        & $\Fs\Fk\alpha$
        & $[(\Fs-1)\Fk + \Fr]\alpha$ \\
        This paper
        & $\Fs\Fk\alpha - \Ir\!\left(\frac{\Fk}{\Fr}-1\right)$
        & $\Fs\Fr\alpha\frac{(\Fs-1)\Fk+\Ir}{(\Fs-1)\Fr+\Ir}$ \\
        \bottomrule
    \end{tabular}
    \vspace{-1em}
\end{table}

\section{Background and related work}
\label{sec:background}

\subsection{Vector codes}
\label{sec:background:vector}

Let $\field$ be the finite field of size $q$ and let $[n,k,\alpha]$ denote a vector code $\Code$ over $\field$, where $\Code \subseteq \field^{\alpha n}$ is defined as an $\field$-linear subspace with dimension $\alpha k$.
We assume $\Code$ has a given basis, which forms the \emph{generator matrix} $\V{G} \in \field^{\alpha k \times \alpha n}$ of a code.
We denote the \emph{encoding function} of $\Code$ as $\Code(\msg):=\msg \V{G}$ for message $\msg \in \field^{\alpha k}$.
If $\V{G}=[\V{I} \mid \V{P}]$, (where $\V{I}$ is the identity matrix) the code is said to be \emph{systematic}.
Let $[n]:=\{1,\ldots,n\}$.
The coordinates of codeword $\V{c} \in \Code$ are called \emph{subsymbols}, and $\V{c}_i = (c_{\alpha(i-1)+1},\allowbreak c_{\alpha(i-1)+2},\ldots,c_{\alpha i})$ is defined as the $i$-th \emph{symbol} of $\V{c}$ ($i \in [n]$).
The code $\Code$ is said to be maximum-distance-separable (MDS) iff for all $\msg \in \field^{\alpha k}$, $\msg$ can be decoded from any $k$ symbols of $\V{c} = \Code(\msg)$.
Codes with $\alpha=1$ are called \emph{scalar}.
The notation $\V{p}[i]$ is used to denote the $i$-th coordinate of vector $\V{p}$.

\subsection{\piggyfw for constructing vector codes}
\label{sec:background:piggyback}

The \emph{\piggyfw}~\cite{rashmi2013piggybacking,rashmi2017piggybacking} is a framework for constructing an $[n,k,\alpha]$ vector code by using $\alpha$ instances of an $[n,k]$ scalar code as a \emph{base code}, and then adding \emph{piggybacks} to certain subsymbols of the code.
The piggybacks are arbitrary functions of data from one instance added to another instance, chosen so as to grant some additional property to the code.
Typically, the piggyback added to instance $i$ is only a function of the data encoded by instances $1,\ldots,i-1$.
This property ensures that the code can be decoded by sequentially decoding instances $1,\ldots,\alpha$ in order, using the data from the decoded instances to subtract the piggybacks.
We will employ piggybacking to design bandwidth efficient convertible codes.
Note that if the base code is MDS, then the constructed vector code with piggybacks is also MDS.

\subsection{Convertible codes}
\label{sec:background:convertible}

Convertible codes~\cite{maturana2020convertible} are erasure codes designed to enable encoded data to undergo efficient conversion.
The objective of conversion is to convert codewords of an initial $[\In,\Ik,\alpha]$ code $\ICode$ into codewords of a final $[\Fn,\Fk,\alpha]$ code $\FCode$ such that the initial and final codewords decode to exactly the same set of data symbols.
Assume, for now, $(\In,\Ik;\Fn,\Fk)$ are given, and $\alpha$ is arbitrary.
In this paper, we focus on the case where both $\ICode$ and $\FCode$ are MDS, and in the so-called \emph{split regime}, where $\Ik=\Fs\Fk$ for some integer $\Fs \geq 2$.
This corresponds to conversions where a single initial codeword of $\ICode$ is split into $\Fs$ final codewords of $\FCode$.
Let $\Ir := (\In - \Ik)$ and $\Fr := (\Fn - \Fk)$.
Let $\Msg := (m_i \in \Field)_{i=1}^{\alpha\Ik}$ be the data to be encoded, and let $\Msg_i := (m_{(i-1)\Fk\alpha+j})_{j=1}^{\alpha\Fk}$ be the data associated with final codeword $i \in [\Fk]$.
A split conversion from initial code $\ICode$ to final $\FCode$ is a procedure that takes $\ICode(\Msg)$ as input and outputs $\{\FCode(\Msg_i) \mid i \in [\Fs]\}$.
Our objective is to design the codes $(\ICode,\FCode)$ and an efficient conversion procedure for the given parameters $(\In, \Ik=\Fs\Fk; \Fn, \Fk)$.

During conversion, there are three types of symbols:
\begin{enumerate*}
    \item
    \emph{unchanged symbols}, which are the initial symbols that are retained in one of the final codewords (this does not require conversion bandwidth because the symbol does not move);
    \item \emph{retired symbols}, which are the remaining initial symbols that are not unchanged; and
    \item \emph{new symbols}, which are the remaining final symbols that are not unchanged.
\end{enumerate*}
During conversion information is downloaded from unchanged and retired symbols, and then used to construct the new symbols.

Convertible codes that have the maximum number of unchanged symbols are called \emph{stable}.
Intuitively, more unchanged symbols imply fewer new symbols, which requires reading and writing less data when creating the new symbols.
Therefore, to simplify our analysis we focus only on stable convertible codes: with $\Fk$ unchanged symbols per final codeword \cite{maturana2020access}.

\subsection{Other related work}
\label{sec:related-work}

Several works have studied problems that can be regarded as special cases of code conversion: \cite{rashmi2011enabling,mousavi2018delayed} studied the bandwidth required by the addition of extra parities to an MDS code ($\Ik=\Fk$ and $\In < \Fn$); \cite{xia2015tale} describes two pairs of non-MDS codes that can be converted back and forth; \cite{su2020local} studies a problem in distributed matrix multiplication where parameters are changed via local re-encoding.
Another related problem is the \emph{scaling problem}~\cite{\scalingall}, which consists of converting each codeword of an $[n,k,\alpha]$ code, into a codeword of an $[n+s, k+s, k\alpha/(k+s)]$ code for given integer $s$.
In other words, the amount of data in each codeword is kept constant, but the data is distributed across a different number of devices.

\section{Conversion bandwidth of the split regime}
\label{sec:split-lower-bound}

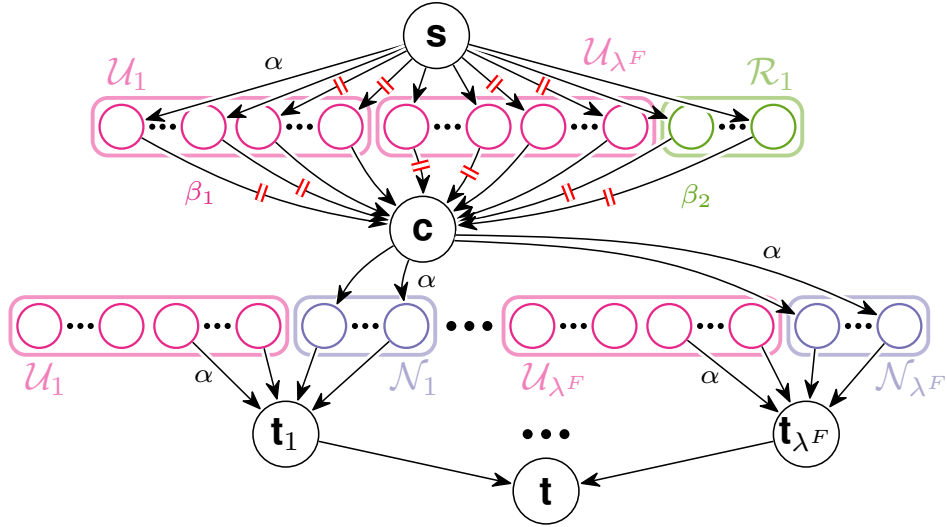
\begin{figure}
    \centering
    \resizebox{.7\linewidth}{!}
    {\begin{tikzpicture}[x=5mm, y=5.2mm]
\colorlet{color1}{Dark2-D}
\colorlet{color2}{Dark2-E}
\colorlet{color3}{Dark2-C}
\colorlet{color4}{Dark2-B}
\colorlet{cut}{red}

\tikzset{snode/.style={draw, circle, inner sep=0pt, minimum size=6mm, font=\sffamily\bfseries}}
\tikzset{enode/.style={draw, circle, inner sep=0pt, minimum size=4mm}}
\tikzset{par/.style={fill=black!30}}
\tikzset{myedge/.style={-{Stealth[round, length=2mm]}, shorten >=.5pt}}
\tikzset{backedge/.style={draw=black!40, fill=black!40, line width=.7pt, dashed}}
\tikzset{around-edge/.style={preaction={-{}, draw=white, line width={#1}}}}

\tikzset{cut-edge/.style={
	decoration={
		markings,
		mark=at position {#1} with
			{\draw [-{}, draw=cut, double distance=1pt, line width=.6pt] (0pt,2.5pt) -- (0pt, -2.5pt); }			
		},
		postaction={decorate},
},
cut-edge/.default={.65}}

\tikzset{ellipsis/.style 2 args={
	draw=white, fill,
	decorate, decoration={
		pre=moveto,
		pre length={#1},
		post=moveto,
		post length={#1},		
		shape backgrounds,
		shape=circle,
		shape size={#2},
		shape evenly spread=3,
	},
},
ellipsis/.default={2pt}{2pt}}

\node [snode] (source-0) at (5.75, -1.1) {$\mbox{s}$};

\node [snode] (sink-0) at (3, -8.1) {$\mbox{t}_1$};
\node [snode] (sink-1) at (12.5, -8.1) {$\mbox{t}_{\lambda^F}$};
\draw [ellipsis={19mm}{3pt}] (sink-0) -- (sink-1);

\node [snode] (sink) at (15.5/2, -9.1) {$\mbox{t}$};

\draw [myedge] (sink-0) edge (sink);
\draw [myedge] (sink-1) edge (sink);

\begin{scope}[shift={(0, -2.7)}]
	\foreach \i/\x in {0/0, 1/5.2} {
        \begin{scope}[every node/.append style={draw=color1, line width=.6pt}]
          \node [enode] (initial-data-\i-0) at (\x, 0) {};
          \node [enode] (initial-data-\i-1) at (\x+1.5, 0) {};
          \node [enode] (initial-data-\i-2) at (\x+2.5, 0) {};
          \node [enode] (initial-data-\i-3) at (\x+4, 0) {};
          \draw [ellipsis] (initial-data-\i-0) -- (initial-data-\i-1);
          \draw [ellipsis] (initial-data-\i-2) -- (initial-data-\i-3);
        \end{scope}
    }
    \begin{scope}[every node/.append style={draw=color2, line width=.6pt}]
      \node [enode] (initial-par-0-0) at (5.2+5.2, 0) {};
      \node [enode] (initial-par-0-1) at (5.2+6.7, 0) {};
      \draw [ellipsis] (initial-par-0-0) -- (initial-par-0-1);
    \end{scope}
\end{scope}

\begin{scope}[shift={(0, -6.2)}]
	\foreach \i/\x in {0/-1.5, 1/7.5} {
    \begin{scope}[every node/.append style={draw=color1, line width=.6pt}]
      \node [enode] (final-data-\i-0) at (\x, 0) {};
      \node [enode] (final-data-\i-1) at (\x+1.5, 0) {};
      \node [enode] (final-data-\i-2) at (\x+2.5, 0) {};
      \node [enode] (final-data-\i-3) at (\x+4, 0) {};
      \draw [ellipsis] (final-data-\i-0) -- (final-data-\i-1);
      \draw [ellipsis] (final-data-\i-2) -- (final-data-\i-3);
    \end{scope}
		
    \begin{scope}[every node/.append style={draw=color3, line width=.6pt}]
      \node [enode] (final-par-\i-0) at (\x+5.2, 0) {};
      \node [enode] (final-par-\i-1) at (\x+6.7, 0) {};
      \draw [ellipsis] (final-par-\i-0) -- (final-par-\i-1);
    \end{scope}
	}
	\draw [ellipsis={2mm}{3pt}] (final-par-0-1) -- (final-data-1-0);
\end{scope}


\node [snode] (center) at (5.5,  -4.5) {c};

\draw [myedge] (initial-data-0-0) edge [bend right=10, around-edge={2pt}, cut-edge=.5] (center);
\draw [myedge] (initial-data-0-1) edge [bend right=10, around-edge={2pt}, cut-edge=.5] (center);
\draw [myedge] (initial-data-0-2) edge [bend right=10, around-edge={2pt}] (center);
\draw [myedge] (initial-data-0-3) edge [bend right=10, around-edge={2pt}] (center);

\draw [myedge] (initial-data-1-0) edge [cut-edge=.38, bend left=10, around-edge={2pt}] (center);
\draw [myedge] (initial-data-1-1) edge [cut-edge=.4, bend left=10, around-edge={2pt}] (center);
\draw [myedge] (initial-data-1-2) edge [bend left=10, around-edge={2pt}] (center);
\draw [myedge] (initial-data-1-3) edge [bend left=10, around-edge={2pt}] (center);

\foreach \j in {0,1} {
	\draw [myedge] (initial-par-0-\j) edge [cut-edge=.5, bend left=10, around-edge={2pt}] (center);
}

\foreach \i in {0, 1} \foreach \j in {0, 1} {
	\ifthenelse{\i = 0}{
		\draw [myedge] (center) edge [bend right=15, around-edge={2pt}] (final-par-\i-\j);
	}{
		\draw (center.-20+10*\j) edge [myedge, bend left=15, out=10, around-edge={2pt}] (final-par-\i-\j);
	}
}

\begin{scope}[every edge/.append style={myedge, around-edge={2pt}}]
  \draw (source-0) edge (initial-data-0-0);
  \draw (source-0) edge (initial-data-0-1);
  \draw (source-0) edge [cut-edge=.52] (initial-data-0-2);
  \draw (source-0) edge [cut-edge=.5] (initial-data-0-3);
  \draw (source-0) edge (initial-par-0-0);
  \draw (source-0) edge (initial-par-0-1);

  \draw (source-0) edge (initial-data-1-0);
  \draw (source-0) edge (initial-data-1-1);
  \draw (source-0) edge [cut-edge=.5] (initial-data-1-2);
  \draw (source-0) edge [cut-edge=.55] (initial-data-1-3);
\end{scope}

\begin{scope}[every edge/.append style={myedge, around-edge={2pt}}]
  \foreach \i in {0, 1} {
    \draw (final-data-\i-2) edge (sink-\i);
    \draw (final-data-\i-3) edge (sink-\i);
    \draw (final-par-\i-0) edge (sink-\i);
    \draw (final-par-\i-1) edge (sink-\i);
  }
\end{scope}

\begin{scope}[
  every node/.append style={font=\scriptsize},
  every edge/.append style={{}-{},draw=none},
  ]
  \draw (source-0)
  edge node [midway, above] {$\alpha$}
  (initial-data-0-0);
  \draw (initial-par-0-1)
  edge [bend left=10] node [xshift=15pt, midway, right, font=\scriptsize\color{color2}] {$\beta_2$}
  (center);
  \draw (initial-data-0-0)
  edge [bend right=10] node [xshift=-8pt, midway, left, font=\scriptsize\color{color1}] {$\beta_1$}
  (center);
  \draw (center)
  edge [bend right=15] node [midway, right] {$\alpha$}
  (final-par-0-1);
  \draw (center)
  edge [bend left=10] node [xshift=20pt, midway, right] {$\alpha$}
  (final-par-1-1);
  \draw (final-data-0-2)
  edge node [midway, left] {$\alpha$}
  (sink-0);
  \draw (final-data-1-2)
  edge node [midway, left] {$\alpha$}
  (sink-1);
\end{scope}

\begin{scope}[
    on background layer,
    nodebox/.style={
      draw={#1!50},
      line width=1pt,
      rounded corners,
      inner sep=1.5pt,
    },
    boxlabel/.style 2 args={
      {#2}=-2pt,
      font=\color{#1!60},
    },
  ]
  \node (unchanged-i-0)
  [
    nodebox=color1,
    fit=(initial-data-0-0)(initial-data-0-3),
  ] {};
  \node at (unchanged-i-0.north west)
    [boxlabel={color1}{above}, xshift=3.5mm]
    {$\mathcal{U}_{1}$};

  \node (unchanged-i-1)
  [
    nodebox=color1,
    fit=(initial-data-1-0)(initial-data-1-3),
  ] {};
  \node at (unchanged-i-1.north east)
    [boxlabel={color1}{above}, xshift=-3.5mm, yshift=1.7mm]
    {$\mathcal{U}_{\lambda^F}$};

  \node (unchanged-f-0)
  [
    nodebox=color1,
    fit=(final-data-0-0)(final-data-0-3),
  ] {};
  \node at (unchanged-f-0.south west)
    [boxlabel={color1}{below}, xshift=3.5mm]
    {$\mathcal{U}_{1}$};

  \node (unchanged-f-1)
  [
    nodebox=color1,
    fit=(final-data-1-0)(final-data-1-3),
  ] {};
  \node at (unchanged-f-1.south west)
    [boxlabel={color1}{below}, xshift=5mm]
    {$\mathcal{U}_{\lambda^F}$};

  \node (retired-0)
  [
    nodebox=color2,
    fit=(initial-par-0-0)(initial-par-0-1),
  ] {};
  \node at (retired-0.north east)
    [boxlabel={color2}{above}, xshift=-3mm]
    {$\mathcal{R}_1$};

  \node (new-0)
  [
    nodebox=color3,
    fit=(final-par-0-0)(final-par-0-1),
  ] {};
  \node at (new-0.south east)
    [boxlabel={color3}{below}, xshift=-2mm]
    {$\mathcal{N}_1$};

  \node (new-1)
  [
    nodebox=color3,
    fit=(final-par-1-0)(final-par-1-1),
  ] {};
  \node at (new-1.south east)
    [boxlabel={color3}{below}, xshift=-1.5mm]
    {$\mathcal{N}_{\lambda^F}$};
\end{scope}

\end{tikzpicture}}
    \caption{%
        Information flow graph of split conversion.
        For clarity, each unchanged symbol is drawn twice, in order to show the initial configuration of the system in the top row of nodes, and the final configuration in the bottom row of nodes.
        The edges with a red mark depict a graph cut.
        }
    \label{fig:split-flow}
\end{figure}

In this section we analyze the conversion bandwidth required by MDS convertible codes in the split regime, i.e., the case where $\Ik = \Fs\Fk$ for some integer $\Fs \geq 2$.

In order to obtain a lower bound on the conversion bandwidth, we model split conversion as an information flow problem.
In this model, we represent the flow of information during conversion as a DAG with edges with variable capacity that represent the transfer of data between nodes. 
Our objective is to set the capacity of edges in a way that minimizes the conversion bandwidth, while ensuring that the flow conditions necessary for conversion are met.

One challenge is that, as we will show, the bound we obtain through information flow is not achievable in general.\footnote{Split conversion corresponds to a \emph{multi-source multicast} problem. In this case (unlike the \emph{single-source} case) the information flow bound is not necessarily tight with respect to coding~\cite{yeung2002multi}.}
This bound is not achievable in general because retired symbols contain data that is associated with more than one final codeword. Thus, in order to make use of these symbols during conversion, we must also download enough data from unchanged symbols to remove the ``interference'' from other final codewords.
To this end, we introduce a conjecture and derive from it a lower bound which, as we show in \cref{sec:split-construction}, is achievable.

\subsection{Information flow}

We model the conversion process using the graph (see \cref{fig:split-flow}) composed by the following nodes:
\begin{itemize}
    \item source $\sourcev$, representing the whole data $\Msg \in \field^{\alpha\Ik}$;
    \item the set $\mathcal{U}_i$ for $i \in [\Fs]$, representing the unchanged symbols of final codeword $i$;
    \item the set $\mathcal{R}$ representing retired symbols;
    \item the set $\mathcal{N}_i$ for $i \in [\Fs]$, representing the new symbols of final codeword $i$;
    \item data collectors $\sinkvs{i}$ for $i \in [\Fs]$ that represent the decoders for each final codeword;
    \item a central node $\centralv$ that computes the new symbols;
    \item a sink $\sinkv$ collecting the data for all final codewords (i.e.\ $\msg$).
\end{itemize}
Let $(u,v,x)$ denote and edge from node $u$ to node $v$ with capacity $x \geq 0$.
Nodes are connected by the following edges:
\begin{itemize}
    \item $\{(\sourcev, x, \alpha) \mid x \in \bigcup_i \mathcal{U}_i \cup \mathcal{R}\}$, representing the data stored in the initial symbols;
    \item $\{(x, \centralv, \beta_1) \mid x \in \bigcup_i \mathcal{U}_i\}$ representing the data downloaded from unchanged symbols;
    \item $\{(x, \centralv, \beta_2) \mid x \in \mathcal{R}\}$, representing the data downloaded from retired symbols;
    \item $\{(\centralv, x, \alpha) \mid x \in \bigcup_i \mathcal{N}_i\}$, representing the data written to the new symbols;
    \item $\{(x, \sinkvs{i}, \alpha) \mid x \in V_i\}$ for $V_j \subseteq \bigcup_i (\mathcal{U}_i \cup \mathcal{N}_i)$ such that $|V_j| = \Fk$ for $j \in [\Fs]$, representing decoding of final codeword $i$;
    \item $\{(\sinkvs{i}, \sinkv, \alpha\Fk) \mid i \in [\Fs]\}$, representing the collection of all the decoded data.
\end{itemize}
In this paper, we focus on stable codes (see \cref{sec:background:convertible}).
Therefore, we have that $|\mathcal{U}_i| = \Fk$, $|\mathcal{R}| = \Ir$, and $|\mathcal{N}_i| = \Fr$ ($i \in [\Fs]$).
The total conversion bandwidth $\gamma$ will be given by the total size of the information communicated between nodes during conversion, which corresponds to the following equation:
\begin{equation}\label{eq:split-bandwidth}
\begin{gathered}
    \gamma := \gamma_R + \gamma_W,\\
    \text{where }
    \gamma_R := \Fs\Fk\beta_1 + \Ir\beta_2
    \text{ and }
    \gamma_W := \Fs\Fr\alpha.
\end{gathered}
\end{equation}
We refer to $\gamma_R$ as the \emph{read conversion bandwidth} and to $\gamma_W$ as the \emph{write conversion bandwidth}.
Our objective is to set $(\beta_1, \beta_2)$ to minimize $\gamma$ while ensuring an information flow of size $\alpha\Ik$ (the size of the data $\msg$) is feasible.
Since $\gamma_W$ is constant with respect to $(\beta_1, \beta_2)$, our analysis will focus on $\gamma_R$.

Note that our model assumes a uniform amount of data downloaded from unchanged symbols and retired symbols.
This is without loss of generality, since any stable convertible code with non-uniform downloads, can be made uniform by repeating the code a sufficient number of times and rotating the assignment of symbols to nodes with each repetition.

Our first lemma expresses the constraint which arises from considering the cut shown in \cref{fig:split-flow}.
\begin{lemma}\label{thm:split-flow}
    For all stable MDS \SplitCode:
    \begin{equation} \label{eq:split-flow}
        \Fs\min\{\Fr,\Fk\}\alpha \leq \Fs\min\{\Fr,\Fk\}\beta_1 + \Ir\beta_2.
    \end{equation}
\end{lemma}
\begin{proof}
    For each $j \in [\Fs]$, consider a sink $\sinkvs{j}$ that connects to all symbols in a final codeword but a set $S_j \subseteq \mathcal{U}_{j}$ of size $\min\{\Fk,\Fr\}$.
    Consider the cut defined by $\{\sourcev\} \cup \bigcup_{j=1}^{\Fs} S_j \cup \mathcal{R}$.
    This cut yields \eqref{eq:split-flow} after simplification.
\end{proof}
Using \eqref{eq:split-bandwidth}, we can show that when $\Fr \geq \Fk$, no savings in conversion bandwidth are possible over the default approach.
\begin{corollary}\label{thm:no-savings}
    When $\Fr \geq \Fk$, we have $\gamma_R \geq \Fs\Fk\alpha$.
    \xqed{\IEEEQED}
\end{corollary}
In other words, the default approach has optimal conversion bandwidth when $\Fr \geq \Fk$.
For this reason, we will only focus on the case $\Fr < \Fk$.

To obtain a lower bound on $\gamma$, we will minimize it subject to \eqref{eq:split-flow} with $\beta_1$ and $\beta_2$ as variables.
\begin{lemma}\label{thm:beta-loose}
    Assume $\Fr < \Fk$.
    Then, the value of $\gamma$ is minimized subject to \eqref{eq:split-flow} when:
    \begin{equation*}
        \beta_1 = \max\left\{1 - \frac{\Ir}{\Fs\Fr},0\right\} \alpha, \qquad
        \beta_2 = \min\left\{1,\frac{\Fs\Fr}{\Ir}\right\} \alpha.
    \end{equation*}
\end{lemma}
\begin{proof}
    As intuition, note that $\beta_2$ offers the better ``bang for the buck'' for satisfying \eqref{eq:split-flow}, because each unit of $\beta_2$ contributes $\Ir$ costing $\Ir$, while each unit of $\beta_1$ contributes $\Fs\Fr$ costing $\Fs\Fk$.
    Thus, in order to satisfy \eqref{eq:split-flow}, it is better to increase $\beta_2$ first, and then increase $\beta_1$ if necessary.
    This approach leads to the proposed solution.
    It is straightforward to check that this solution satisfies the Karush-Kuhn-Tucker (KKT) conditions, and is thus an optimal solution.
\end{proof}
By replacing into \eqref{eq:split-bandwidth}, we obtain the following lower bound.
\begin{theorem}\label{thm:split-bound-loose}
    For all stable MDS \SplitCode:
    \begin{equation*}\label{eq:split-bound-loose}
        \gamma_R \geq
        \begin{cases}
            \Fs\!\Fk\alpha - \Ir\alpha\max\left\{\frac{\Fk}{\Fr}-1, 0\right\} &\text{if $\Ir \leq \Fs\!\Fr$,}\\
            \Fs\min\{\Fr, \Fk\}\alpha &\text{otherwise.}\\
        \end{cases}%
    \end{equation*}
\end{theorem}
\begin{proof}
    Follows from \cref{thm:beta-loose} and case analysis.
\end{proof}
This bound shows that there is potential for conversion bandwidth savings when $\Fk > \Fr$, because the bound is strictly lower than the default approach ($\Fs\Fk\alpha$) in this region.
Unfortunately, this bound is not always achievable, as we see next.

For example, suppose we have have a stable convertible code with $\Fk > \Fr$, $\Ir = \Fs\Fr$ and that we set $\beta_1 = 0$ and $\beta_2 = \alpha$.
This assignment satisfies \cref{eq:split-bound-loose} (and it is easy to check that it leads to a feasible flow in \cref{fig:split-flow}).
However, as shown by previous work on access cost of conversion~\cite{maturana2020access}, it is not possible to perform conversion in this case by accessing fewer than $(\Fs - 1)\Fk + \Fr$ symbols.
Furthermore, it can be shown that any assignment that makes $\beta_1 > 0$ necessarily leads to a higher conversion bandwidth than the lower bound of \cref{eq:split-bound-loose}.
The fundamental problem in this case is that to create new symbols for a particular final codeword we need to remove the interference from all other final codewords.
This is not possible if the conversion procedure does not access a sufficient number of symbols.

For this reason, we introduce the following conjecture, which lower bounds the amount of data that needs to be downloaded from unchanged symbols based on the above intuition.
\begin{conjecture}\label{thm:conjecture}
    In the information flow model presented in this section, for all stable MDS \SplitCode we must have:
    \begin{equation}\label{eq:conjecture}
        \Fs\beta_1 \geq (\Fs - 1)\beta_2.
    \end{equation}
\end{conjecture}
We incorporate this constraint into the minimization of $\gamma$ and obtain a different solution, which limits the amount of data downloaded from retired symbols when $\Ir > \Fr$.
\begin{lemma}\label{thm:beta-tight}
    Assume $\Fr < \Fk$.
    Then, the minimum value of $\gamma$ subject to \eqref{eq:split-flow} and \eqref{eq:conjecture} is achieved by \cref{thm:beta-loose} when $\Ir < \Fr$, and otherwise by:
    \begin{equation*}
        \beta_1 = \frac{(\Fs-1)\Fr\alpha}{(\Fs-1)\Fr + \Ir},
        \qquad
        \beta_2 = \frac{\Fs\Fr\alpha}{(\Fs-1)\Fr+\Ir}.
    \end{equation*}
\end{lemma}
\begin{proof}
    As in the case of \cref{thm:beta-loose}, it is intuitively better to increase $\beta_2$ rather than $\beta_1$.
    However, \eqref{eq:conjecture} gives an upper bound on $\beta_2$ in terms of $\beta_1$.
    Therefore, we set $\beta_2 = \min\left\{\alpha, \frac{\Fs}{\Fs-1}\beta_1\right\}$.
    We then replace $\beta_2$ in \eqref{eq:split-flow} and set $\beta_1$ in order to satisfy the inequality.
    When $\Ir < \Fr$, one can check that \eqref{eq:conjecture} is not tight, and thus we obtain the same values that \cref{thm:beta-loose}.
    Otherwise, we obtain the stated values of $\beta_1$ and $\beta_2$.
    It is straightforward to check that this solution satisfies the KKT conditions, and is thus an optimal solution.
\end{proof}
By replacing back into \eqref{eq:split-bandwidth}, we obtain the following lower bound based on \cref{thm:conjecture}.

\begin{theorem}\label{thm:split-bound-tight}
    If \cref{thm:conjecture} holds, then for all \SplitCode with $\Ir \geq \Fr$ and $\Fr \leq \Fk$:
    \begin{equation*}\label{eq:split-bound-tight}
        \gamma_R \geq \Fs\Fr\alpha \frac{(\Fs-1)\Fk+\Ir}{(\Fs-1)\Fr+\Ir}.
    \end{equation*}
\end{theorem}
\begin{proof}
    Follows from \cref{thm:beta-tight}.
\end{proof}
As we shall see in \cref{sec:split-construction}, the proposed constructions achieve the combination of the lower bounds of \cref{thm:split-bound-loose,thm:split-bound-tight}.
Thus, we finish this section by comparing the conversion bandwidth of our approach with that of the default approach and existing convertible codes optimized for access cost~\cite{maturana2020access}.
Since in all approaches the write conversion bandwidth is equal ($\Fs\Fr\alpha$), we focus on the read conversion bandwidth.
\Cref{tab:bw-comparison} includes the expressions for the read conversion bandwidth of different approaches.
\Cref{fig:split-bandwidth-plot} plots the lower bounds on read conversion bandwidth relative to the default approach for some example parameters.
These results show that our approach can achieve significant savings in conversion bandwidth with respect to the default approach and access-optimal convertible codes.

\section{Explicit constructions}
\label{sec:split-construction}

In this section, we present constructions for convertible codes in the split regime that optimize for conversion bandwidth.
The constructions employ the Piggybacking framework~\cite{rashmi2017piggybacking}.

\begin{theorem}\label{thm:split-construction}
    The constructions presented in this section achieve the optimal conversion bandwidth when $\Ir \leq \Fr$.
    Furthermore, they achieve the optimal conversion bandwidth when $\Ir > \Fr$ if \cref{thm:conjecture} is true.
\end{theorem}
\begin{proof}
    Follows directly from the design and description of the constructions below.
\end{proof}
These construction require less conversion bandwidth than the default approach \emph{and} the access optimal approach (regardless of \cref{thm:conjecture}) as long as $\Fr < \Fk$ (\cref{thm:no-savings}).
We begin by describing the base code used in both constructions, and then present the piggybacking constructions for the cases $\Ir > \Fr$ and $\Ir \leq \Fr$, respectively.

\subsection{The base code}
\label{sec:split-construction:base}
We utilize an $[\In, \Ik]$ systematic code with a Vandermonde matrix with evaluation points $(\xi_1,\ldots,\xi_{\Ir})$ as the parity matrix.
A code of this form is guaranteed to be MDS when choosing $\xi_t$ ($t \in [\Ir]$) and field size as specified by the general construction in \cite{maturana2020convertible}.
Nonetheless, in practice it is often possible to search for $\xi_t$ that generate an MDS code over a given finite field.
Let $\V{h}_t := (h^{(t)}_1,\ldots,h^{(t)}_{\Ik})^T = (1,\xi_t,\ldots,\xi_t^{\Ik-1})^T$ be parity encoding vector $t \in [\Ir]$ of the base code.
In our construction, we use the property that $(h^{(t)}_1,\ldots,h^{(t)}_{\Fk}) = \xi_t^{-(i-1)\Fk}(h^{(t)}_{(i-1)\Fk+1},\ldots,h^{(t)}_{i\Fk})$ for all $t \in [\Ir]$ and $i \in [\Fs]$.

\subsection{Piggybacking construction (case \texorpdfstring{$\Ir > \Fr$}{rI > rF})}
We now describe the construction (assuming $\Fr < \Fk$).
Recall that during conversion, we download $\beta_1$ from each unchanged symbol, and $\beta_2$ from each retired symbol, which are set as discussed in \cref{sec:split-lower-bound}.
If we set the set the size of each symbol as $\alpha := ((\Fs-1)\Fr + \Ir)$, then $\beta_1 := (\Fs-1)\Fr$ and $\beta_2 := \Fs\Fr$.
For simplicity, we divide $\alpha$ into blocks: for a given $\ell \in [\alpha]$ we define $(\ell_1, \ell_2)$ as follows.
\begin{equation*}
    (\ell_1,\ell_2) :=
    \begin{cases}
        \left(\left\lceil\frac{\ell}{\Fr}\right\rceil,\, (\ell-1 \bmod \Fr) + 1\right) & \text{if $\ell \leq \Fs\Fr$,}\\
        (\Fs + 1,\, \ell - \Fs\Fr) & \text{otherwise.}
    \end{cases}
\end{equation*}
To describe the encoding vectors of our code, we decompose each encoding vector of the base code into $\Fs$ vectors of length $\Fk$, corresponding to the data associated with each final codeword.
Then, we represent each of these vector in the $\alpha\Ik$-dimensional space corresponding to the whole data $\msg$ (by filling the additional dimensions with zeros).
Specifically, we define $\V{p}^{(i)}_{t,\ell} \in \Field^{\alpha\Ik}$ as the column vector such that $\msg\V{p}^{(i)}_{t,\ell}$ corresponds to the encoding of the data under the base code for parity $t \in [\Ir]$, final codeword $i \in [\Fs]$, and instance $\ell \in [\alpha]$.
We achieve this by setting $\V{p}^{(i)}_{t,\ell}[(i-1)\Fk\alpha+(j-1)\alpha+\ell] := \V{h}_t[(i-1)\Fk + j]$ for $j \in [\Fk]$ and 0 everywhere else. 

We specify how to construct $\V{q}^I_{t,\ell} \in \Field^{\alpha\Ik}$, which is the encoding vector for instance $\ell \in [\alpha]$ of parity $t \in [\Ir]$ of the initial codeword, and $\V{q}^{F(i)}_{t,\ell} \in \Field^{\alpha\Ik}$ which is the encoding vector for instance $\ell \in [\alpha]$ of parity $t \in [\Fr]$ of final codeword $i \in [\Fs]$.
The construction is designed so that the final codewords are all encoded under the same final code.
\Cref{fig:split-construction-down-red} shows a diagram for this construction.
The construction has three important elements:
\begin{enumerate}
    \item 
    \emph{Permutation}:
    In the initial code, the first $\Fs$ blocks of the data symbols associated with final codeword $i$ are circularly shifted to the right $i - 1$ times (denoted with letters \textsf{\textbf{A}}-\textsf{\textbf{C}}).
    This reordering is logical (no data is moved) and used for describing the code only.
    \item
    \emph{Projection}:
    For parities 1 through $\Fr$ (\textsf{\textbf{P}} blocks), we use the base code without modification to encode each data column.
    During conversion, we download blocks $\{2,\ldots,\Fs\}$ from each data symbol (blocks \textsf{\textbf{B}} and \textsf{\textbf{C}}) and subtract their interference from the corresponding parity symbols to obtain the first block of each final codeword (\textsf{\textbf{P}}) blocks).
    \item
    \emph{Piggybacks}:
    For parities $(\Fr + 1)$ through $\Ir$ (\textsf{\textbf{Q}} blocks), we use the base code and add piggybacks to block $\ell_1 \in [\Fs]$ that contain the subsymbols of block $(\Fs + 1)$ of final codeword $\ell_1$ (transposed).
    During conversion, we recover the piggybacks by using the downloaded data (blocks \textsf{\textbf{B}} and \textsf{\textbf{C}}).
    Note that the piggybacks will still have extra data remaining from the unaccessed block (\textsf{\textbf{A}}).
    However, the final code can still be sequentially decoded (the same way that codes in the piggyback framework are decoded).
\end{enumerate}
The remaining parity subsymbols are generated from the accessed data blocks (\textsf{\textbf{B}} and \textsf{\textbf{C}}).
Finally, parity symbol $t \in [\Fr]$ in final codeword $i \in [\Fs]$ is scaled by $\xi_t^{-(i-1)\Fk}$ to ensure that all final codewords are encoded by the same final code (as described in \cref{sec:split-construction:base}).
Let $\overrightarrow{\ell}\!(i) := ((\ell_1-i \bmod \Fs)\Fk + \ell_2)$ be the instance index after permutation.
Then, the encoding vectors for the initial and final codes are defined as:
\begin{gather*}
    \V{q}^I_{t,\ell} \!:=\!
    \begin{cases}
        \sum_{i=1}^{\Fs} \V{p}^{(i)}_{t,\overrightarrow{\ell}\!(i)} & \text{if } t \leq \Fr\!,\, \ell_1 \leq \Fs\!,\\
        \sum_{i=1}^{\Fs} \V{p}^{(i)}_{t,\overrightarrow{\ell}\!(i)} \!+ 
        \tikzmark{pbs}
        \V{p}^{(\ell_1)}_{\ell_2,(\Fs-1)\Fr\!+t}
        \tikzmark{pbe}
        & \text{if } t > \Fr\!,\, \ell_1 \leq \Fs\!,\\
        \sum_{i=1}^{\Fs} \V{p}^{(i)}_{t,\ell} & \text{otherwise}.\\
    \end{cases}\\
    \V{q}^{F(i)}_{t,\ell} :=
    \begin{cases}
        \xi_t^{-(i-1)\Fk}
        \V{p}^{(i)}_{t,\ell} & \text{if } \ell_1 \leq \Fs,\\
        \tikzmark{sfs}
        \xi_t^{-(i-1)\Fk}
        \tikzmark{sfe}
        (\V{p}^{(i)}_{t,\ell} + 
        \tikzmark{eds}
        \V{p}^{(i)}_{\Fr+\ell_2,t}
        \tikzmark{ede}
        ) & \text{otherwise}.
    \end{cases}
    \tikz[remember picture] \draw[overlay,decorate,decoration={brace,raise=2.1mm,mirror},gray] (pic cs:pbs) -- (pic cs:pbe) node[midway,below=2.3mm] {\scriptsize piggyback};
    \tikz[remember picture] \draw[overlay,decorate,decoration={brace,raise=2mm,mirror},gray] (pic cs:sfs) -- (pic cs:sfe) node[midway,below=2.1mm] {\scriptsize scaling factor};
    \tikz[remember picture] \draw[overlay,decorate,decoration={brace,raise=2mm,mirror},gray] (pic cs:eds) -- (pic cs:ede) node[midway,below=2.1mm] {\scriptsize extra data};
\end{gather*}
\vspace{.4em}

Notice that during conversion we only need to download $\beta_1 = (\Fs-1)\Fr$ subsymbols from each unchanged symbol and $\beta_2 = \Fs\Fr$ subsymbols from each retired symbol, out of the $\alpha = ((\Fs-1)\Fr+\Ir)$ subsymbols in each symbol.
Therefore, this construction achieves the conversion bandwidth bound of \cref{thm:split-bound-tight}.
Furthermore, the constructed code is MDS because it uses the piggyback framework and the base code is MDS.

\subsection{Piggybacking construction (case \texorpdfstring{$\Ir \leq \Fr$}{rI <= rF})}

The construction in the case when $\Ir \leq \Fr$ is similar.
In this case, we set $\alpha := \Fs\Fr$, and $\beta_1 := (\Fs\Fr - \Ir)$ and $\beta_2 := \Fs\Fr$.
We divide each symbol evenly into $\Fs$ blocks of $\Fr$ columns.
Thus, for a given $\ell \in [\alpha]$ we define $(\ell_1, \ell_2)$ as follows.
\begin{align*}
    \ell_1 &:= \left\lceil\frac{\ell}{\Fr}\right\rceil,\\
    \ell_2 &:= (\ell-1 \bmod \Fr) + 1.
\end{align*}

Now we describe the construction.
\Cref{fig:split-construction-up-red} shows a diagram for this construction.
\begin{enumerate}
    \item
    \emph{Permutation}:
    In the initial code, the $\Fs$ blocks of the data symbols associated with final codeword $i$ are circularly shifted to the right $i - 1$ times (denoted with letters \textsf{\textbf{A}}-\textsf{\textbf{C}}).
    This reordering is logical (no data is moved) and used for describing the code only.
    \item
    \emph{Projection}:
    For the first $\Ir$ columns of each block, we use the base code without modification to encode each data column.
    During conversion, we download the data in the first $\Ir$ columns of blocks $\{2,\ldots,\Fs\}$ from each data symbol (blocks \textsf{\textbf{B}} and \textsf{\textbf{C}}) and subtract their interference from the corresponding parity symbols to obtain the first $\Ir$ columns of the first block of parities $1$ through $\Ir$ in each final codeword (\textsf{\textbf{P}}) blocks).
    \item
    \emph{Piggybacks}:
    For columns $(\Ir + 1)$ through $\Fr$ of each block (\textsf{\textbf{Q}} blocks), we use the base code and add piggybacks that contain the subsymbols of the first $\Ir$ columns of the first block of parities $(\Ir + 1)$ through $\Fr$ in each final codeword $\ell_1$ (transposed).
    During conversion, we download all the data in the corresponding columns (blocks \textsf{\textbf{A'}}, \textsf{\textbf{B}} and \textsf{\textbf{C}}) and recover the piggybacks.
\end{enumerate}
The remaining parity subsymbols are generated from the accessed data blocks (\textsf{\textbf{A'}}, \textsf{\textbf{B}}, and \textsf{\textbf{C}}).
Finally, parity symbol $t \in [\Fr]$ is scaled by $\xi_t^{-(i-1)\Fk}$.
Let $\overrightarrow{\ell}\!(i)$ and $\V{p}^{(i)}_{t,\ell}$ be defined as before (in the case $\Ir > \Fr$).
Then, the encoding vectors for the initial and final codes are defined as:
\begin{gather*}
    \V{q}^I_{t,\ell} \!:=\!
    \begin{cases}
        \sum_{i=1}^{\Fs} \V{p}^{(i)}_{t,\overrightarrow{\ell}\!(i)} & \text{if } \ell_2 \leq \Ir\!,\\
        \sum_{i=1}^{\Fs} \V{p}^{(i)}_{t,\overrightarrow{\ell}\!(i)} \!+ 
        \tikzmark{pbs2}
        \V{p}^{(\ell_1)}_{\ell_2,t}
        \tikzmark{pbe2}
        & \text{otherwise.}\\
    \end{cases}\\
    \\
    \V{q}^{F(i)}_{t,\ell} :=
        \tikzmark{sfs2}
        \xi_t^{-(i-1)\Fk}
        \tikzmark{sfe2}
        \V{p}^{(i)}_{t,\ell}
    \tikz[remember picture] \draw[overlay,decorate,decoration={brace,raise=2.1mm,mirror},gray] (pic cs:pbs2) -- (pic cs:pbe2) node[midway,below=2.3mm] {\scriptsize piggyback};
    \tikz[remember picture] \draw[overlay,decorate,decoration={brace,raise=2mm,mirror},gray] (pic cs:sfs2) -- (pic cs:sfe2) node[midway,below=2.1mm] {\scriptsize scaling factor};\\
\end{gather*}
Notice that during conversion we only need to download $\beta_1 = (\Fs\Fr-\Ir)$ subsymbols from each unchanged symbol and $\beta_2 = \Fs\Fr$ subsymbols from each retired symbol, out of the $\alpha = \Fs\Fr$ subsymbols in each symbol.
Therefore, this construction achieves the conversion bandwidth bound of \cref{thm:split-bound-tight}.
Furthermore, the constructed code is MDS because it uses the piggyback framework and the base code is MDS.

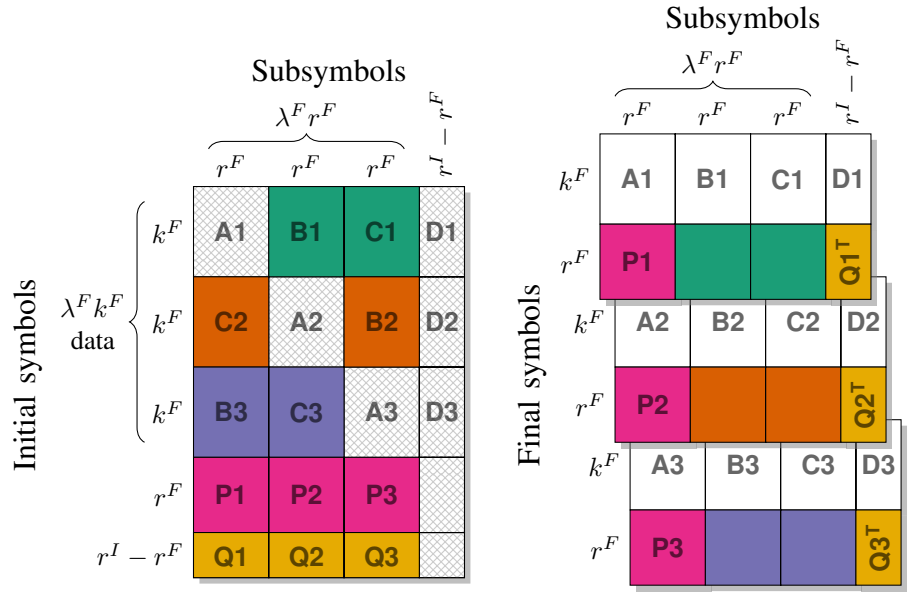
\begin{figure}
    \centering
    \resizebox{.7\linewidth}{!}{%
    \newcommand{\kfs}{1.2}
\newcommand{\ris}{1.6}
\newcommand{\rfs}{1}
\newcommand{\rwidth}{\ris+2*\rfs}
\newcommand{\rheight}{3*\kfs-\ris}
\begin{tikzpicture}
    \pgfdeclarelayer{bg1}
    \pgfdeclarelayer{bg2}
    \pgfsetlayers{bg2,bg1,main}
    \tikzset{ltrlabel/.style={opacity=.6, font=\bfseries\sffamily}}
    
    \draw[fill=white,drop shadow]
    (3*\rfs,0) rectangle ++(\ris-\rfs, -\kfs)
    (3*\rfs,-\kfs) rectangle ++(\ris-\rfs, -\kfs)
    (3*\rfs,-2*\kfs) rectangle ++(\ris-\rfs, -\kfs)
    (3*\rfs,-3*\kfs) rectangle ++(\ris-\rfs, -\rfs)
    (3*\rfs,-3*\kfs-\rfs) rectangle ++(\ris-\rfs, -\ris+\rfs)
    (0,-3*\kfs-\rfs) rectangle ++(\rfs,-\ris+\rfs)
    (\rfs,-3*\kfs-\rfs) rectangle ++(\rfs,-\ris+\rfs)
    (2*\rfs,-3*\kfs-\rfs) rectangle ++(\rfs,-\ris+\rfs)
    ;
    
    \draw [pattern=crosshatch, pattern color=black!20]
    (3*\rfs,0) rectangle ++(\ris-\rfs, -\kfs)
    (3*\rfs,-\kfs) rectangle ++(\ris-\rfs, -\kfs)
    (3*\rfs,-2*\kfs) rectangle ++(\ris-\rfs, -\kfs)
    (3*\rfs,-3*\kfs) rectangle ++(\ris-\rfs, -\rfs)
    (3*\rfs,-3*\kfs-\rfs) rectangle ++(\ris-\rfs, -\ris+\rfs)
    (0,0) rectangle ++(\rfs, -\kfs)
    (\rfs,-\kfs) rectangle ++(\rfs, -\kfs)
    (2*\rfs,-2*\kfs) rectangle ++(\rfs, -\kfs)
    ;

    \draw [fill=Dark2-A]
    (\rfs,0) rectangle ++(\rfs,-\kfs)
    (2*\rfs,0) rectangle ++(\rfs,-\kfs)
    ;
    \draw [fill=Dark2-B]
    (0,-\kfs) rectangle ++(\rfs,-\kfs)
    (2*\rfs,-\kfs) rectangle ++(\rfs,-\kfs)
    ;
    \draw [fill=Dark2-C]
    (0,-2*\kfs) rectangle ++(\rfs,-\kfs)
    (\rfs,-2*\kfs) rectangle ++(\rfs,-\kfs)
    ;
    \draw [fill=Dark2-D]
    (0,-3*\kfs) rectangle ++(\rfs,-\rfs)
    (\rfs,-3*\kfs) rectangle ++(\rfs,-\rfs)
    (2*\rfs,-3*\kfs) rectangle ++(\rfs,-\rfs)
    ;
    \draw [fill=Dark2-F]
    (0,-3*\kfs-\rfs) rectangle ++(\rfs,-\ris+\rfs)
    (\rfs,-3*\kfs-\rfs) rectangle ++(\rfs,-\ris+\rfs)
    (2*\rfs,-3*\kfs-\rfs) rectangle ++(\rfs,-\ris+\rfs)
    ;

    \node at (0,-\kfs/2) [left] {$k^F$};
    \node at (0,-\kfs*3/2) [left] {$k^F$};
    \node at (0,-\kfs*5/2) [left] {$k^F$};
    \node at (0,-\kfs*3-\rfs/2) [left] {$r^F$};
    \node at (0,-\kfs*3-\rfs-\ris/2+\rfs/2) [left] {$r^I-r^F$};

    \node at (\rfs/2,0) [above] {$r^F$};
    \node at (\rfs*3/2,0) [above] {$r^F$};
    \node at (\rfs*5/2,0) [above] {$r^F$};
    \node at (3*\rfs+\ris/2-\rfs/2,0) [right,rotate=90] {$r^I - r^F$};

    \draw [decorate, decoration={brace, raise=6.5mm, amplitude=2mm}]
    (0, -3*\kfs+.2) -- (0,-.2) node[midway, left=7.8mm, align=center] {$\lambda^F k^F$\\data};
    \draw [decorate, decoration={brace, raise=5mm, amplitude=2mm}]
    (.2, 0) -- (3*\rfs-.2,0) node[midway, above=6.8mm] {$\lambda^F r^F$};
    
    \node at (-2.2,-2.5) [rotate=90] {\large Initial symbols};
    \node at (4.5,-2.5) [rotate=90] {\large Final symbols};
    
    \node at (1.8,1.5) {\large Subsymbols};
    \node at (7.3,2.2) {\large Subsymbols};

    \begin{scope}[xshift=5.4cm, yshift=-.5cm]
        \node at (0,\kfs/2) [left] {$k^F$};
        \node at (0,-\rfs/2) [left] {$r^F$};
        \node at (\rfs/2,\kfs) [above] {$r^F$};
        \node at (\rfs*3/2,\kfs) [above] {$r^F$};
        \node at (\rfs*5/2,\kfs) [above] {$r^F$};
        \node at (3*\rfs+\ris/2-\rfs/2,\kfs) [right,rotate=90] {$r^I - r^F$};
    
        \draw [decorate, decoration={brace, raise=5mm, amplitude=2mm}]
        (.2,\kfs) -- (3*\rfs-.2,\kfs) node[midway, above=6.8mm] {$\lambda^F r^F$};

        \draw[fill=white,drop shadow]
        (0,\kfs) rectangle ++(\rfs,-\kfs)
        (\rfs,\kfs) rectangle ++(\rfs,-\kfs)
        (2*\rfs,\kfs) rectangle ++(\rfs,-\kfs)
        (3*\rfs,\kfs) rectangle ++(\ris-\rfs,-\kfs)
        ;
        \draw [fill=Dark2-D, drop shadow]
        (0,0) rectangle ++(\rfs,-\rfs)
        ;
        \draw [fill=Dark2-A, drop shadow]
        (\rfs,0) rectangle ++(\rfs,-\rfs)
        (2*\rfs,0) rectangle ++(\rfs,-\rfs)
        ;
        \draw [fill=Dark2-F, drop shadow]
        (3*\rfs,0) rectangle ++(\ris-\rfs, -\rfs)
        ;
    \end{scope}
    \begin{pgfonlayer}{bg1}
    \begin{scope}[xshift=5.6cm, yshift=-2.4cm]
        \node at (0,\kfs/2) [left] {$k^F$};
        \node at (0,-\rfs/2) [left] {$r^F$};

        \draw[fill=white, drop shadow]
        (0,\kfs) rectangle ++(\rfs,-\kfs)
        (\rfs,\kfs) rectangle ++(\rfs,-\kfs)
        (2*\rfs,\kfs) rectangle ++(\rfs,-\kfs)
        (3*\rfs,\kfs) rectangle ++(\ris-\rfs,-\kfs)
        ;
        \draw [fill=Dark2-D, drop shadow]
        (0,0) rectangle ++(\rfs,-\rfs)
        ;
        \draw [fill=Dark2-B, drop shadow]
        (\rfs,0) rectangle ++(\rfs,-\rfs)
        (2*\rfs,0) rectangle ++(\rfs,-\rfs)
        ;
        \draw [fill=Dark2-F, drop shadow]
        (3*\rfs,0) rectangle ++(\ris-\rfs, -\rfs)
        ;
    \end{scope}
    \end{pgfonlayer}
    \begin{pgfonlayer}{bg2}
    \begin{scope}[xshift=5.8cm, yshift=-4.3cm]
        \node at (0,\kfs/2) [left] {$k^F$};
        \node at (0,-\rfs/2) [left] {$r^F$};

        \draw [fill=white, drop shadow]
        (0,\kfs) rectangle ++(\rfs,-\kfs)
        (\rfs,\kfs) rectangle ++(\rfs,-\kfs)
        (2*\rfs,\kfs) rectangle ++(\rfs,-\kfs)
        (3*\rfs,\kfs) rectangle ++(\ris-\rfs,-\kfs)
        ;
        \draw [fill=Dark2-D, drop shadow]
        (0,0) rectangle ++(\rfs,-\rfs)
        ;
        \draw [fill=Dark2-C, drop shadow]
        (2*\rfs,0) rectangle ++(\rfs,-\rfs)
        (\rfs,0) rectangle ++(\rfs,-\rfs)
        ;
        \draw [fill=Dark2-F, drop shadow]
        (3*\rfs,0) rectangle ++(\ris-\rfs, -\rfs)
        ;
    \end{scope}
    \end{pgfonlayer}
    
    \foreach \i
    [evaluate=\i as \in using {-.6 -1.2*\i},
     evaluate=\i as \il using {int(\i+1)}]
    in {0,1,2} {
        \foreach \j/\ltr
        [evaluate=\j as \jn using {mod(.5+\i+\j, 3)}]
        in {0/A,1/B,2/C} {
            \node [ltrlabel] at (\jn, \in) {\ltr\il};
        }
        \node [ltrlabel] at (3.3, \in) {D\il};
    }
    \foreach \i
    [evaluate=\i as \in using {.1 -1.9*\i},
     evaluate=\i as \il using {int(\i + 1)}]
    in {0,1,2} {
        \foreach \j/\ltr
        [evaluate=\j as \jn using {5.9+\j}]
        in {0/A,1/B,2/C} {
            \node [ltrlabel] at (\jn+.2*\i, \in) {\ltr\il};
        }
        \node [ltrlabel] at (8.7+.2*\i, \in) {D\il};
    }
    \foreach \i
    [evaluate=\i as \in using {.5+\i},
     evaluate=\i as \il using {int(\i + 1)}]
    in {0,1,2} {
        \node [ltrlabel] at (\in, -4.1) {P\il};
        \node [ltrlabel] at (\in, -4.93) {Q\il};
    }
    \foreach \i
    [evaluate=\i as \in using {.5+\i},
     evaluate=\i as \il using {int(\i + 1)}]
    in {0,1,2} {
        \node [ltrlabel] at (5.9+.2*\i, -1-1.9*\i) {P\il};
        \node [ltrlabel,rotate=90] at (8.7+.2*\i, -1-1.9*\i) {Q\il\textsuperscript{T}};
    }
\end{tikzpicture}
    }
    \caption{Diagram of convertible code construction in the split regime when $\Ir > \Fr$ and $\Fs = 3$.}
    \label{fig:split-construction-down-red}
\end{figure}
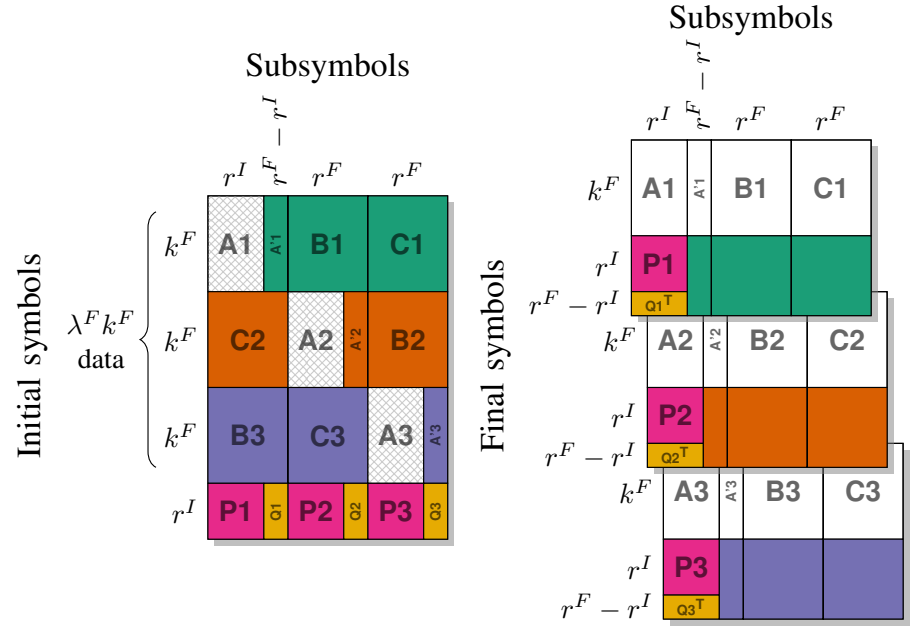
\begin{figure}
    \centering
    \resizebox{.7\linewidth}{!}{%
    \newcommand{\kfs}{1.2}
\newcommand{\ris}{.7}
\newcommand{\rfs}{1}
\newcommand{\rwidth}{\ris+2*\rfs}
\newcommand{\rheight}{3*\kfs-\ris}
\begin{tikzpicture}
    \pgfdeclarelayer{bg1}
    \pgfdeclarelayer{bg2}
    \pgfsetlayers{bg2,bg1,main}
    \tikzset{ltrlabel/.style={opacity=.6, font=\bfseries\sffamily}}
    
    \draw[fill=white,drop shadow]
    (0,0) rectangle ++(3*\rfs,-3*\kfs-\ris)
    ;
    
    \draw [fill=gray, pattern=crosshatch, pattern color=black!20]
    (0,0) rectangle ++(\ris,-\kfs)
    (\rfs,-\kfs) rectangle ++(\ris,-\kfs)
    (2*\rfs,-2*\kfs) rectangle ++(\ris,-\kfs)
    ;
    \draw [fill=Dark2-A]
    (\ris,0) rectangle ++(\rfs-\ris,-\kfs)
    (\rfs,0) rectangle ++(\rfs,-\kfs)
    (2*\rfs,0) rectangle ++(\rfs,-\kfs)
    ;
    \draw [fill=Dark2-B]
    (0,-\kfs) rectangle ++(\rfs,-\kfs)
    (\rfs+\ris,-\kfs) rectangle ++(\rfs-\ris,-\kfs)
    (2*\rfs,-\kfs) rectangle ++(\rfs,-\kfs)
    ;
    \draw [fill=Dark2-C]
    (0,-2*\kfs) rectangle ++(\rfs,-\kfs)
    (\rfs,-2*\kfs) rectangle ++(\rfs,-\kfs)
    (2*\rfs+\ris,-2*\kfs) rectangle ++(\rfs-\ris,-\kfs)
    ;
    \draw [fill=Dark2-D]
    (0,-3*\kfs) rectangle ++(\ris,-\ris)
    (\rfs,-3*\kfs) rectangle ++(\ris,-\ris)
    (2*\rfs,-3*\kfs) rectangle ++(\ris,-\ris)
    ;
    \draw [fill=Dark2-F]
    (\ris,-3*\kfs) rectangle ++(\rfs-\ris,-\ris)
    (\rfs+\ris,-3*\kfs) rectangle ++(\rfs-\ris,-\ris)
    (2*\rfs+\ris,-3*\kfs) rectangle ++(\rfs-\ris,-\ris)
    ;
    
    \node at (0,-\kfs/2) [left] {$k^F$};
    \node at (0,-\kfs*3/2) [left] {$k^F$};
    \node at (0,-\kfs*5/2) [left] {$k^F$};
    \node at (0,-\kfs*3-\ris/2) [left] {$r^I$};

    \node at (\ris/2,0) [above] {$r^I$};
    \node at (\ris+\rfs/2-\ris/2,0) [right,rotate=90] {$r^F - r^I$};
    \node at (\rfs*3/2,0) [above] {$r^F$};
    \node at (\rfs*5/2,0) [above] {$r^F$};

    \draw [decorate, decoration={brace, raise=6.5mm, amplitude=2mm}]
    (0, -3*\kfs+.2) -- (0,-.2) node[midway, left=7.8mm, align=center] {$\lambda^F k^F$\\data};

    \node at (-2.2,-2) [rotate=90] {\large Initial symbols};
    \node at (3.6,-2) [rotate=90] {\large Final symbols};
    
    \node at (1.5,1.6) {\large Subsymbols};
    \node at (6.8,2.2) {\large Subsymbols};

    \begin{scope}[xshift=5.3cm, yshift=-.5cm]
        \node at (0,\kfs/2) [left] {$k^F$};
        \node at (0,-\ris/2) [left] {$r^I$};
        \node at (0,-\rfs/2-\ris/2) [left] {$r^F-r^I$};
        
        \node at (\ris/2,\kfs) [above] {$r^I$};
        \node at (\ris+\rfs/2-\ris/2,\kfs) [right,rotate=90] {$r^F - r^I$};
        \node at (\rfs*3/2,\kfs) [above] {$r^F$};
        \node at (\rfs*5/2,\kfs) [above] {$r^F$};

        \draw[fill=white, drop shadow]
        (0,\kfs) rectangle ++(3*\rfs,-\kfs-\rfs)
        ;
        
        \draw
        (0,\kfs) rectangle ++(\ris,-\kfs)
        (\ris,\kfs) rectangle ++(\rfs-\ris,-\kfs)
        (\rfs,\kfs) rectangle ++(\rfs,-\kfs)
        (2*\rfs,\kfs) rectangle ++(\rfs,-\kfs)
        ;
        \draw [fill=Dark2-D]
        (0,0) rectangle ++(\ris,-\ris)
        ;
        \draw [fill=Dark2-F]
        (0,-\ris) rectangle ++(\ris,-\rfs+\ris)
        ;
        \draw [fill=Dark2-A]
        (\ris,0) rectangle ++(\rfs-\ris,-\rfs)
        (\rfs,0) rectangle ++(\rfs,-\rfs)
        (2*\rfs,0) rectangle ++(\rfs,-\rfs)
        ;
    \end{scope}
    \begin{pgfonlayer}{bg1}
    \begin{scope}[xshift=5.5cm, yshift=-2.4cm]
        \node at (0,\kfs/2) [left] {$k^F$};
        \node at (0,-\ris/2) [left] {$r^I$};
        \node at (0,-\rfs/2-\ris/2) [left] {$r^F-r^I$};

        \draw[fill=white, drop shadow]
        (0,\kfs) rectangle ++(3*\rfs,-\kfs-\rfs)
        ;
        \draw
        (0,\kfs) rectangle ++(\ris,-\kfs)
        (\ris,\kfs) rectangle ++(\rfs-\ris,-\kfs)
        (\rfs,\kfs) rectangle ++(\rfs,-\kfs)
        (2*\rfs,\kfs) rectangle ++(\rfs,-\kfs)
        ;
        \draw [fill=Dark2-D]
        (0,0) rectangle ++(\ris,-\ris)
        ;
        \draw [fill=Dark2-F]
        (0,-\ris) rectangle ++(\ris,-\rfs+\ris)
        ;
        \draw [fill=Dark2-B]
        (\ris,0) rectangle ++(\rfs-\ris,-\rfs)
        (\rfs,0) rectangle ++(\rfs,-\rfs)
        (2*\rfs,0) rectangle ++(\rfs,-\rfs)
        ;
    \end{scope}
    \end{pgfonlayer}
    \begin{pgfonlayer}{bg2}
    \begin{scope}[xshift=5.7cm, yshift=-4.3cm]
        \node at (0,\kfs/2) [left] {$k^F$};
        \node at (0,-\ris/2) [left] {$r^I$};
        \node at (0,-\rfs/2-\ris/2) [left] {$r^F-r^I$};

        \draw[fill=white, drop shadow]
        (0,\kfs) rectangle ++(3*\rfs,-\kfs-\rfs)
        ;
        \draw
        (0,\kfs) rectangle ++(\ris,-\kfs)
        (\ris,\kfs) rectangle ++(\rfs-\ris,-\kfs)
        (\rfs,\kfs) rectangle ++(\rfs,-\kfs)
        (2*\rfs,\kfs) rectangle ++(\rfs,-\kfs)
        ;
        \draw [fill=Dark2-D]
        (0,0) rectangle ++(\ris,-\ris)
        ;
        \draw [fill=Dark2-F]
        (0,-\ris) rectangle ++(\ris,-\rfs+\ris)
        ;
        \draw [fill=Dark2-C]
        (\ris,0) rectangle ++(\rfs-\ris,-\rfs)
        (\rfs,0) rectangle ++(\rfs,-\rfs)
        (2*\rfs,0) rectangle ++(\rfs,-\rfs)
        ;
    \end{scope}
    \end{pgfonlayer}

    \foreach \i
    [evaluate=\i as \in using {-.6 -1.2*\i},
     evaluate=\i as \il using {int(\i+1)}]
    in {0,1,2} {
        \foreach \j/\ltr/\adj
        [evaluate=\j as \jn using {mod(.5+\i+\j, 3)}]
        in {0/A/-.15,1/B/0,2/C/0} {
            \node [ltrlabel] at (\jn+\adj, \in) {\ltr\il};
        }
    }
    \foreach \i
    [evaluate=\i as \in using {.1 -1.9*\i},
     evaluate=\i as \il using {int(\i + 1)}]
    in {0,1,2} {
        \foreach \j/\ltr/\adj
        [evaluate=\j as \jn using {5.8+\j}]
        in {0/A/-.15,1/B/0,2/C/0} {
            \node [ltrlabel] at (\jn+\adj+.2*\i, \in) {\ltr\il};
        }
    }
    \foreach \i
    [evaluate=\i as \in using {.35+\i},
     evaluate=\i as \il using {int(\i + 1)}]
    in {0,1,2} {
        \node [ltrlabel] at (\in, -3.95) {P\il};
        \node [ltrlabel,rotate=90] at (\in+.5, -3.95) {\tiny Q\il};
        \node [ltrlabel,rotate=90] at (\in+.5, -0.6-1.2*\i) {\tiny A'\il};
    }
    \foreach \i
    [evaluate=\i as \in using {.5+\i},
     evaluate=\i as \il using {int(\i + 1)}]
    in {0,1,2} {
        \node [ltrlabel] at (5.65+.2*\i, -.85-1.9*\i) {P\il};
        \node [ltrlabel] at (5.65+.2*\i, -1.35-1.9*\i) {\tiny Q\il\textsuperscript{T}};
        \node [ltrlabel,rotate=90] at (6.15+.2*\i, .1-1.9*\i) {\tiny A'\il};
    }
\end{tikzpicture}
    }
    \caption{Diagram of convertible code construction in the split regime when $\Ir \leq \Fr$ and $\Fs=3$.}
    \label{fig:split-construction-up-red}
\end{figure}

\section{Conclusion}
This work shows that it is possible to achieve conversion in the split regime with significantly less conversion bandwidth than existing approaches.
Furthermore, this work shows that this reduction can be achieved by using the piggybacking framework, which results in constructions that are relatively simple and have low subpacketization.

\bibliographystyle{IEEEtran}
\bibliography{IEEEabrv,main}

\end{document}